\documentclass[twocolumn,superscriptaddress,floatfix]{revtex4-1}

\usepackage[dvipdfmx]{graphicx}
\usepackage{dcolumn}
\usepackage{bm}
\usepackage{multirow}

\usepackage{color}
\usepackage{ascmac}

\begin{document}

\title{
Single-site orthogonalization for first-principles computations of exchange coupling constants
}
\author{Asako Terasawa}
\email{terasawa.a.aa@m.titech.ac.jp}
\author{Sonju Kou}
\affiliation{
Department of Materials Science and Engineering, Tokyo Institute of Technology,
J1-3, Nagatsuta-cho 4259, Midori-ku, Yokohama 226-8502, Japan}
\author{Taisuke Ozaki}
\affiliation{
Institute for Solid State Physics, The University of Tokyo,
Kashiwanoha 5-1-5, Kashiwa 277-8581, Japan}
\author{Yoshihiro Gohda}
\email{gohda.y.ab@m.titech.ac.jp}
\affiliation{
Department of Materials Science and Engineering, Tokyo Institute of Technology,
J1-3, Nagatsuta-cho 4259, Midori-ku, Yokohama 226-8502, Japan}

\begin{abstract}
For accurate first-principles computations of exchange coupling constants $J_{ij}$ by the Liechtenstein method with localized basis sets, we developed a scheme using  the single-site orthogonalization (SO). In contrast to the non-orthogonal (NO) scheme, where the basis set is used to compute $J_{ij}$ without modification, and the L\"owdin orthogonalization (LO) scheme, the SO scheme 
exhibits 
much less dependence of $J_{ij}$ on the choice of the basis set. The SO scheme achieves convergence of $J_{ij}$ for bcc Fe, hcp Co, and fcc Ni with an increase in the number of the basis set, while the NO and LO schemes result 
in the fluctuation depending on the basis set.
This improvement by the SO scheme is attributed to the removal of orbital overlaps with avoiding ill-defined single-site effective potentials. We further improve the SO scheme by introducing appropriate spin population, so that the SO with spin-population scaling (SOS) scheme 
can provide 
converged Curie temperatures for transition metals. Moreover, negative values of $J_{ij}$ for dhcp Nd and rhombohedral Sm obtained by the SOS scheme can coincide with the experimentally-found magnetic order that cannot be reproduced by positive sets of $J_{ij}$.
\end{abstract}

\maketitle

\section{Introduction}

Improving the performance 
of permanent magnets is one of the 
most pressing technological requirements 
from industry for achieving 
high 
energy-efficient society. The key 
to this 
lies in understanding the microstructures of permanent magnets, because the microstructure properties determine the movement of the internal magnetic domain walls \cite{S_Sugimoto_2011, K_Hono_2012, S_Hirosawa_2017, S_Li_2002, W_F_Li_2009, T_H_Kim_2012, H_Sepehri_Amin_2012, U_M_R_Seelam_2016}.
The structural complexities of permanent magnets, however, often prevents us from reaching a simple understanding of their magnetic properties. For example, crystallinity and compositions of grain boundary phases vary depending on their local environments \cite{T_T_Sasaki_2016, X_D_Xu_2018}.

Recently, first-principles calculation techniques have been employed to investigate the magnetic properties of permanent magnets \cite{T_Fukushima_2015, B_Balasubramanian_2016, A_Saengdeejing_2016, Y_Tatetsu_2016, Z_Torbatian_2016, N_Umetsu_2016, T_Fukazawa_2017, H_Akai_2018, Y_Gohda_2018, C_E_Patrick_2018, Y_Tatetsu_2018, C_E_Patrick_2019, A_M_Schonhobel_2019, A_L_Tedstone_2019, T_Fukazawa_2019}.
Because of the 
multiple phases and types of internal interfaces 
in permanent magnets, 
many problems remain unsolved.
In particular, 
the exchange couplings between the various phases are important in achieving high-performance permanent magnets \cite{H_Sepehri_Amin_2012}.

\begin{figure}[hbtp]
\begin{center}
\includegraphics[width=70mm]{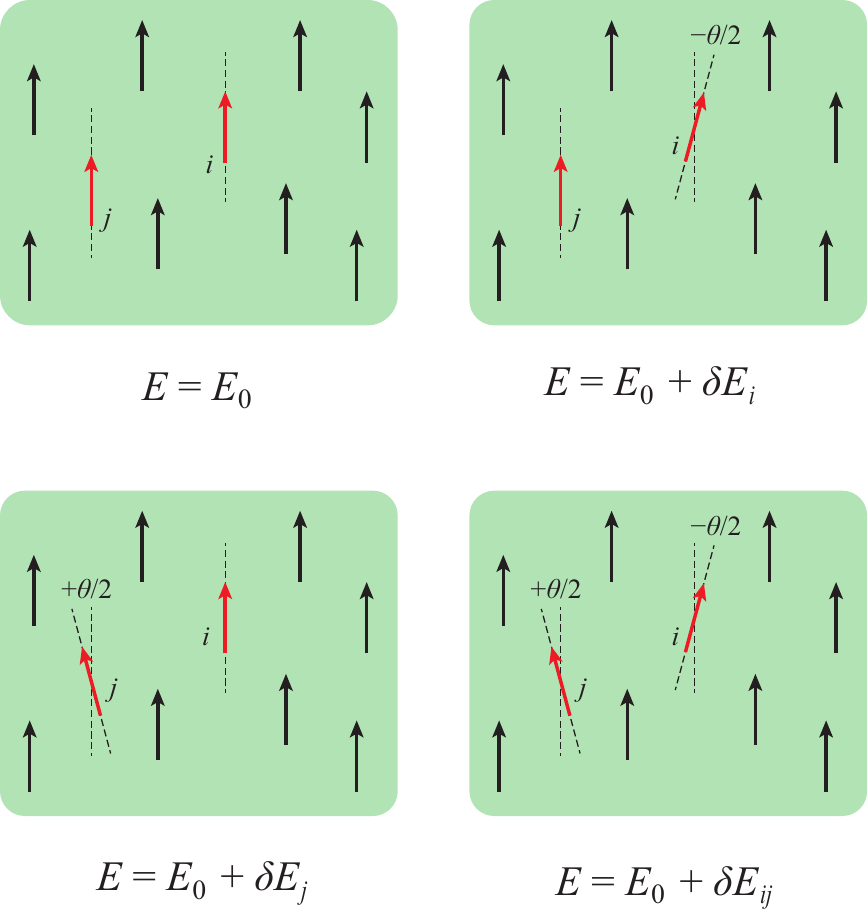}\\
\end{center}
\caption{
A schematic of infinitesimal spin rotations in Liechtenstein method. 
\label{fig:Liechtenstein}
}
\end{figure}

The idea of extracting the exchange coupling constants from ground state calculations was first established by Oguchi \textit{et al.}~\cite{T_Oguchi_1983} and further developed by Liechtenstein \textit{et al.}~\cite{A_I_Liechtenstein_1987}.
The magnetic force theorem, also known as the Liechtenstein method, is a powerful tool for computing of the exchange coupling constants in the classical Heisenberg model.
The Hamiltonian of the classical Heisenberg model can be written as in the following equation:
\begin{equation}
E=-\sum_{i}\sum_{j\ne i}J_{ij}\mathbf{s}_{i}\cdot\mathbf{s}_{j},
\end{equation}
where $\mathbf{s}_{i}$ is the unit vector along the spin direction of site $i$.
In the Liechtenstein method, one aims to extract the exchange coupling constant $J_{ij}$ by calculating the difference in energy response to infinitesimal
rotations of effective potential 
at sites $i$ and $j$ shown schematically in Fig.~\ref{fig:Liechtenstein}:
\begin{eqnarray}
\delta' E_{ij}\equiv
\delta E_{ij}-\delta E_{i}-\delta E_{i}
  =J_{ij}\frac{\theta^2}{2}.
\label{eq:deltadashEij_Jij}
\end{eqnarray}
Making use of the second perturbation theory, 
the exchange coupling constant can be written as
\begin{eqnarray}
J_{ij}
&=
 &\frac{1}{4\pi}\int d\varepsilon\,f_{\mathrm{F}}(\varepsilon)\,
\mathrm{ImTr}\left[
    \hat{P}_{i}
    \hat{G}_{\uparrow}(\varepsilon)
    \hat{P}_{j}
    \hat{G}_{\downarrow}(\varepsilon)
  \right],
\label{eq:Liech_Green}
\end{eqnarray}
where $\hat{G}_{\sigma}(\varepsilon)$ is the retarded Green's function of the spin $\sigma$ for collinear states, and
\[
\hat{P}_{i}
\equiv\hat{V}_{i,\uparrow}-\hat{V}_{i,\downarrow}
\]
is the difference in electronic potential at site $i$.

Since its establishment, 
there have been many works on improving Liechtenstein methods within the KKR Green's function formalism 
\cite{H_Shiba_1971,H_Akai_1977,H_Akai_1982,S_Lounis_2005,H_Ebert_2011,D_S_G_Bauer_2013}, 
and applying the Liechtenstein method to other electronic calculation schemes, such as the linear muffin-tin orbital method \cite{M_Pajda_2001,I_Turek_2003,A_Szilva_2017,Y_O_Kvashnin_2015,S_Frota-Pessoa_2000,H_Wang_2010} and the linear combination of atomic orbitals (LCAO) approximation \cite{M_J_Han_2004,H_Yoon_2018,A_Terasawa_2019}.
A cumbersome problem arises here: the exchange coupling constants in 
the previous studies often fluctuate on the order of meV, which results in fluctuations of a few hundred Kelvins in the  Curie temperature \cite{M_Pajda_2001,S_Frota-Pessoa_2000,Y_O_Kvashnin_2015,H_Wang_2010,H_Yoon_2018,A_Terasawa_2019}.
The dependence of the $J_{ij}$ fluctuation on the computational scheme has been examined by Kvashnin \textit{et al.}\ \cite{Y_O_Kvashnin_2015}.
They compared the $J_{ij}$ calculated with two definitions of basis functions for single atomic sites, namely a simple integration cutoff outside the radius of the muffin-tin potential was adopted in one definition, and L\"owdin orthogonalized (LO) overlapping basis functions in the other definition.
They found a deviation between the $J_{ij}$ 
calculated 
with the two definitions, and suggested that the deviation 
originated 
from the deviation in the single-site electron populations in the two definitions.
Their conclusion 
highlights the difficulty in determining 
well-defined values of $J_{ij}$, especially in the presence of overlap of atomic orbitals.
This problem is also related to the problem of itinerancy, because wide-ranged basis functions are necessary to 
represent itinerant states accurately.

In this paper, we report a new scheme 
for calculating the
exchange coupling constants in first-principles calculations within 
the 
LCAO approximation.
We examine the calculated $J_{ij}$ values of bcc Fe for different choices of basis sets in detail.
We 
find 
that 
the 
matrix representation of the Hamiltonian using non-orthogonal (NO) atomic orbitals is unsuitable 
for representing 
the single-site potential in the presence of large overlap, and that the computational results vary 
with 
the choice of basis set.
To solve this problem, we introduce a single-site orthogonalization (SO) scheme 
for representing an effective
single-site potential within the LCAO approximation. 
The calculated $J_{ij}$ values 
decrease 
slightly with increasing numbers of basis functions, while 
the 
results 
from 
the NO and LO schemes 
fluctuate significantly with  
the number of basis functions. 
To overcome the slight decrease in $J_{ij}$, we introduce 
a 
spin population scaling in the SO scheme, namely 
the 
single-site orthogonalization with spin-population scaling (SOS).
Using the SOS scheme, we 
successfully obtain 
converged $J_{ij}$ 
curves as functions 
of the atomic distance $r_{ij}$ when the number of basis functions 
is increased.
We also 
apply 
the SOS scheme to 
calculate 
the $J_{ij}$ of various systems 
and their 
transition temperatures within the mean field approximation.
It is then possible to obtain converged Curie temperatures for bcc Fe, hcp Co and fcc Ni. 
For dhcp Nd and rhombohedral Sm, we obtained weak negative $J_{ij}$ curves.
Our results corresponds to the experimental reports of spiral and complicated magnetic orders of Nd and Sm \cite{Coeybook} indicating antiferromagnetic exchange couplings.

\section{Single-site potential and orthogonalization scheme}\label{sec:ortho}

To formulate the Liechtenstein method within the LCAO approximation, it is necessary to first define the single-site potential in the presence of overlap of atomic orbitals.
The simplest approximation is just to apply $i$- and $j$-th diagonal elements of the total Hamiltonian:
\begin{eqnarray}
P_{ii}
&\equiv
 & H_{ii,\uparrow}-H_{ii,\downarrow},\ 
\\
H_{ij,\sigma}
&\equiv
 &\langle i|\hat{H}_\sigma| j\rangle
\end{eqnarray}
to generate the single-site potential operator, where $| i\rangle$ and $| j\rangle$ represent the non-orthogonal atomic orbitals for the sites $i$ and $j$.
For simplicity, we assume here a single orbital for each atomic site.
The corresponding operator notation of $i$-th single-site potential for the above approximation can be written as
\begin{eqnarray}
\hat{P}_{i}^{\mathrm{(NO)}}
&\equiv
 &\sum_{j,j'}|j\rangle\big[\mathbf{S}^{-1}\big]_{ji}P_{ii}\big[\mathbf{S}^{-1}\big]_{ij'}\langle j'|,
\label{eq:op_P_i}
\end{eqnarray}
where the notation (NO) represents that the operator is generated straightforwardly from the matrix representation by non-orthogonal atomic orbitals.
For the definition (\ref{eq:op_P_i}), it is necessary to define the overlap matrix $\mathbf{S}$ for non-orthogonal basis set $\{| j\rangle\}$:
\begin{eqnarray}
\langle i|j\rangle
&=
 &1
\\
\langle i|j\rangle
&=
 &\big[\mathbf{S}\big]_{ij}
  \ (i\ne j),
\end{eqnarray}
where $\displaystyle{\big[ \mathbf{A} \big]_{ij}}$ means the $(i,j)$ element of matrix $\mathbf{A}$.

The physical meaning of $\hat{P}_{i}^{\mathrm{(NO)}}$ is however ambiguous in the presence of large overlap of atomic orbitals belonging to different sites.
When the off-diagonal elements of $\mathbf{S}^{-1}$ are large, many orbitals corresponding to other sites are involved into Eq.~(\ref{eq:op_P_i}), and $\hat{P}_{i}^{\mathrm{(NO)}}$ is no longer single-site like.

To avoid the problem of ambiguity caused by overlap of atomic orbitals, 
orthogonalization schemes are often adopted.
One of the most straightforward orthogonalization schemes is L\"owdin orthogonalization (LO):
\begin{eqnarray}
| \widetilde{i}\rangle
&\equiv
 &\sum_{j}|j\rangle
 \big[ \mathbf{S}^{-1/2} \big]_{ji}.
\end{eqnarray}
We can easily prove that the L\"owdin orbitals belonging to different sites are orthogonal to one another.
With the LO basis set, we define the effective single-site potential by LO scheme as:
\begin{eqnarray}
\hat{P}_{i}^{\mathrm{(LO)}}
&\equiv
 &|\widetilde{i}\rangle P^{\mathrm{(LO)}}_{ii}\langle \widetilde{i}|
\\
P_{ii}^{\mathrm{(LO)}}
&\equiv
 &\langle \widetilde{i}|\hat{H}_{\uparrow}|\widetilde{i}\rangle
  -\langle \widetilde{i}|\hat{H}_{\downarrow}|\widetilde{i}\rangle.
\end{eqnarray}

\begin{figure*}[tp]
\begin{center}
\includegraphics[width=170mm]{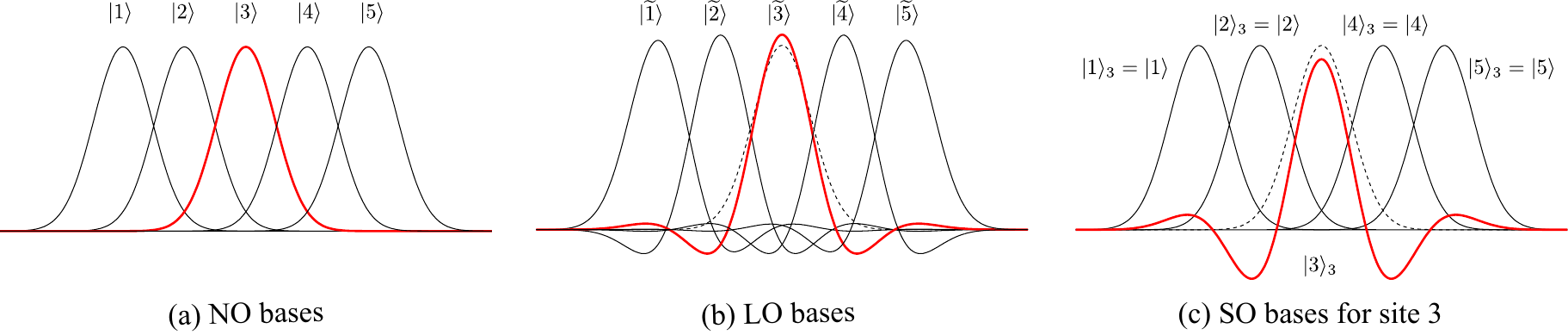}\\
\end{center}
\caption{
A non-orthogonal basis set and its corresponding orthogonalized basis sets, described by one-dimensional Gaussian functions and their linear combinations.
In each panel, the red solid line represents the basis belonging to site 3, and black solid lines represent the basis functions belonging to other sites.
In panels (b) and (c), the dotted black line represents the original NO basis at site 3.
(Color online)
\label{fig:ortho}
}
\end{figure*}

In addition to the LO basis set,
we 
also consider 
single-site orthogonalized (SO) orbitals in this study.
In the SO scheme, the orbital belonging to a particular site in the basis set is 
made 
orthogonal to other orbitals, whereas other pairs of atomic orbitals remain 
unchanged.
To define the SO 
bases 
for site $i$, we first define the 
neighbor set 
$\overline{i}$ for site $i$ as:
\begin{equation}
\overline{i}\equiv\big\{ j\,\big|\,j\ne i,\ 
\langle i|j\rangle
\ne 0 \big\}.
\end{equation}
In other words, the neighbor set $\overline{i}$ of site $i$ is defined as the set of sites that has nonzero overlap with site $i$, but excludes site $i$ itself.
It is then possible to define the partial overlap matrix $\mathbf{S}_{\overline{i},i}$ between the site $i$ and the sites in the neighbor set $\overline{i}$ as
\begin{equation}
\big[\mathbf{S}_{\overline{i},i}\big]_{ji}
\equiv 
\langle j|i\rangle,
\,j\in\overline{i}
\end{equation}
and the partial overlap matrix $\mathbf{S}_{\overline{i},\overline{i}}$ between sites in the neighbor set $\overline{i}$ as
\begin{equation}
\big[\mathbf{S}_{\overline{i},\overline{i}}\big]_{jj'}
\equiv 
\langle j|j'\rangle,\,j,j'\in\overline{i}.
\end{equation}
Here, the matrix $\mathbf{S}_{\overline{i},i}$ has a dimension of $N(\overline{i})\times 1$ and the matrix $\mathbf{S}_{\overline{i},\overline{i}}$ has a dimension of $N(\overline{i})\times N(\overline{i})$, where $N(\overline{i})$ is the number of neighboring sites to site $i$.
Given the partial matrices $\mathbf{S}_{\overline{i},\overline{i}}$ and $\mathbf{S}_{\overline{i},i}$, 
the SO bases that isolates site $i$ 
are
\begin{eqnarray}
|i \rangle_{i}
&\equiv
 &|i\rangle
  -\sum_{j\in \overline{i}}| j \rangle
  \big[ \mathbf{S}_{\overline{i},\overline{i}}^{-1}\mathbf{S}_{\overline{i},i}\big]_{ji}
\label{eq:SO_1}
\\
|j \rangle_{i}
&\equiv
 &|j\rangle,\ j \ne i.
\label{eq:SO_2}
\end{eqnarray}
From the definitions, we can easily prove that
\[
\langle i |_{i}|j \rangle_{i}
=0,
\, j\ne i,
\]
while any other combinations of SO bases are left non-orthogonal.

For a simple example, we examine a five-site system with NO bases described by one-dimensional Gaussian functions as shown in Fig.~\ref{fig:ortho}(a), and the corresponding LO and SO basis functions as 
shown
in Figs.~\ref{fig:ortho}(b) and (c), respectively.
It can be seen 
that the LO basis functions have similar structures to the NO basis functions around their 
maxima 
and small damped oscillations at their tails.
In contrast, 
the SO basis functions oscillate more prominently around their respective sites of focus,
as shown in Fig.~\ref{fig:ortho}(c).
This oscillation cancels the overlaps with the other orbitals which are left unchanged from the original NO bases.

Defining the SO basis set, it is possible to define the effective single-site potential as
\begin{eqnarray}
\hat{P}^{(\mathrm{SO})}_{i}
 &\equiv 
  &| i \rangle_{i}
  \mathcal{S}_{ii}^{-1}
\nonumber\\
&&\times\left(
  \langle i|_{i}\hat{H}_{\uparrow}|i\rangle_{i}
  -\langle i|_{i}\hat{H}_{\downarrow}|i\rangle_{i}
  \right)
\nonumber\\
&&\times\mathcal{S}_{ii}^{-1}
  \langle i|_{i}
\\
\mathcal{S}_{ii}
&\equiv
 &\langle i|_{i}|i\rangle_{i}
  =1-\mathbf{S}_{i,\overline{i}}\mathbf{S}_{\overline{i},\overline{i}}^{-1}\mathbf{S}_{\overline{i},i}.
\end{eqnarray}
Applying the above definition to 
Eq.~(\ref{eq:Liech_Green}),
we obtain the exchange coupling constant based on the SO basis as
\begin{eqnarray}
J^{\mathrm{(SO)}}_{ij}
&=
 &\frac{1}{4\pi}
  \int_{-\infty}^{\infty} d
  \varepsilon
  \,f\left(\beta(
  \varepsilon
  -\varepsilon_{\mathrm{F}})\right)
\nonumber \\
&&\times\,\mathrm{Im}\left\{
  {P}^{\mathrm{(SO)}}_{ii}G_{ij,
  \uparrow
  }(
  \varepsilon
  )
  {P}^{\mathrm{(SO)}}_{jj}G_{ji,
  \downarrow
  }(
  \varepsilon
  )
  \right\},
\label{eq:Jij_SO}
\\
P^{\mathrm{(SO)}}_{ii}
&\equiv
 &\langle i|\hat{P}^{\mathrm{(SO)}}_{i}|i\rangle
  =\langle i|_{i}\hat{H}_{\uparrow}|i\rangle_{i}-\langle i|_{i}\hat{H}_{\downarrow}|i\rangle_{i}
\nonumber \\
&=
 &P_{ii}
  -\mathbf{P}_{i,\overline{i}}\mathbf{S}_{\overline{i},\overline{i}}^{-1}\mathbf{S}_{\overline{i},i}
  -\mathbf{S}_{i,\overline{i}}\mathbf{S}_{\overline{i},\overline{i}}^{-1}\mathbf{P}_{\overline{i},i}
\nonumber \\
&&+\mathbf{S}_{i,\overline{i}}\mathbf{S}_{\overline{i},\overline{i}}^{-1}
   \mathbf{P}_{\overline{i},\overline{i}}\mathbf{S}_{\overline{i},\overline{i}}^{-1}\mathbf{S}_{\overline{i},i},
\label{eq:P_SO}
\end{eqnarray}
where
\begin{eqnarray}
\big[\mathbf{P}_{\overline{i},i}\big]_{ji}
&\equiv 
 & 
  H_{ji,\uparrow}-H_{ji,\downarrow},
  \quad j\in\overline{i}
\\
\big[\mathbf{P}_{\overline{i},\overline{i}}\big]_{jj'}
&\equiv 
 & 
  H_{jj',\uparrow}-H_{jj',\downarrow},
  \quad j,j'\in\overline{i}
\\
G_{ij,\uparrow}(\varepsilon)
&\equiv
 &\sum_{n}\frac{C_{i,n\uparrow}C^{*}_{j,n\uparrow}}{\varepsilon+i\eta-\varepsilon_{n\uparrow}}
  \label{eq:g_up_2}
\\
G_{ji,\downarrow}(\varepsilon)
&\equiv
 &\sum_{n}\frac{C_{j,n'\downarrow}C^{*}_{i,n'\downarrow}}{\varepsilon+i\eta-\varepsilon_{n'\downarrow}},
\label{eq:g_down_2}
\end{eqnarray}
and $\varepsilon_{n\sigma}$ and $\mathbf{C}_{n\uparrow}$ are the eigenvalues and eigenvectors of the LCAO Hamiltonian.

\begin{figure*}[hbtp]
\begin{center}
\includegraphics[width=150mm]{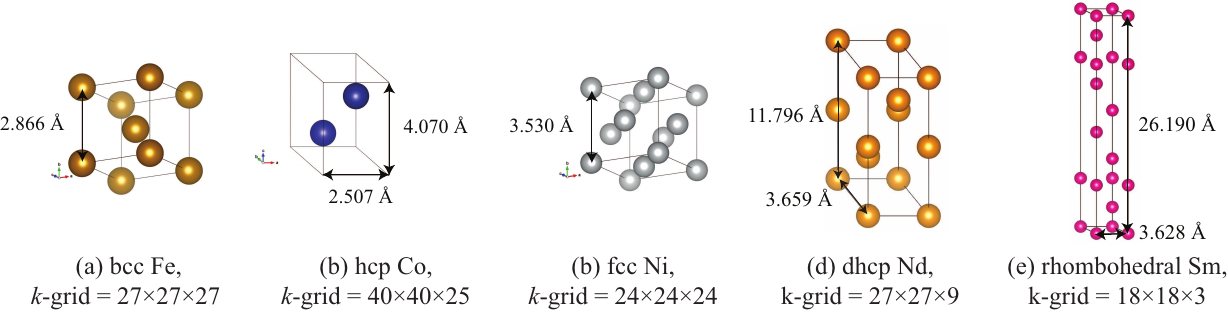}\\
\end{center}
\caption{Computational models of unit cells for transition metals and rare earth metals examined in this study. The notation $k$-grid =$a\times b\times c$ in each figure corresponds to the number of reciprocal space grids in first-principles calculations.
(Color online)
\label{fig:fig_model}
}
\end{figure*}

We can also expand the definitions of LO and SO bases to periodic systems with multi-orbital atoms.
For such systems, 
the
LO bases are written as
\begin{eqnarray}
&&
|\widetilde{\mathbf{R},i,\mu}\rangle
\nonumber \\
&&\equiv
  \sum_{\mathbf{k}}\sum_{j,\nu'}
  | \mathbf{k},j,\nu' \rangle
  \big[\mathbf{S}^{-1/2}\big]_{\mathbf{k}j\nu',\mathbf{R}i\mu}
  e^{-i\mathbf{k}\cdot\mathbf{R}/2},
\end{eqnarray}
and 
the
SO bases are written as
\begin{eqnarray}
&&
|\mathbf{R},i,\mu\rangle_{\mathbf{R},i}
\nonumber \\
&&\equiv |\mathbf{R},i,\mu\rangle
  -\!\!\!\!\!\!
  \sum_{\mathbf{R}'j\in\overline{\mathbf{R}i}}\sum_{\nu}
  | \mathbf{R}',j,\nu \rangle
\nonumber \\
&&\qquad \qquad \qquad \qquad \ \times
  \big[
  \mathbf{S}_{\overline{\mathbf{R}i},\overline{\mathbf{R}i}}^{-1}
  \mathbf{S}_{\overline{\mathbf{R}i},\mathbf{R}i}\big]_{\mathbf{R}'j\nu,\mathbf{R}i\mu}
\label{eq:SO_long_1}
\end{eqnarray}
\begin{eqnarray}
&&
|\mathbf{R}',j,\nu\rangle_{\mathbf{R},i}
  \equiv|\mathbf{R}',j,\nu\rangle,
\,(\mathbf{R}',j) \ne (\mathbf{R},i).
\label{eq:SO_long_2}
\end{eqnarray}
Here, $|\mathbf{R},i,\mu\rangle$ is the $\mu$-th atomic orbital belonging to atom $i$ of cell $\mathbf{R}$, and $|\mathbf{k},i,\mu\rangle$ is the Bloch orbital corresponding to the atomic orbitals of atom $i$.
In Sec.~\ref{sec:results}, we compare the $J_{ij}$ values calculated by the
three
schemes for various systems.

\section{Computational models and methods}

Figure \ref{fig:fig_model} shows the computational models examined in this study,
namely
(a) bcc Fe, (b) hcp Co, (c) fcc Ni, (d) dhcp Nd, and (e) rhombohedral Sm.
The numbers of reciprocal space grid
points
are determined dependent on the systems sizes, and they are 
three
as $k$-grid =$a\times b\times c$ in Fig.~\ref{fig:fig_model}. 

For 
the
first-principles calculations based on
the
LCPAO approximation, we performed the density functional calculations using the OpenMX code \cite{T_Ozaki_2003} (Here PAO in LCPAO means pseudo atomic orbitals, which are basically the same as atomic orbitals except for the exact finite cutoffs).
For the exchange correlation functional, we adopted the Perdew-Burke-Ernzerhof exchange-correlation functional \cite{GGA-PBE} within the generalized gradient approximation (GGA-PBE).
For the pseudo atomic orbitals, the cutoff radii were set to 6.0 Bohr for Fe, Co, and Ni, and to 10 Bohr for Sm and Nd. 
In the spin-dependent SCF calculations, we assumed ferromagnetic configuration of the spin populations for all the systems examined.
We used an electronic temperature of 300 K, and the convergence criterion for the total energy was set as $1.0\times10^{-6}$ Ha.

Our implementation of
the
Liechtenstein formula was based on the finite pole approximation of the Fermi function \cite{T_Ozaki_2007} 
in which
the energy integration over the real axis in Eq.~(\ref{eq:Liech_Green}) was substituted by summation over
a
finite number of poles of 
the
approximated Fermi function.
The implementation is reported 
in detail 
in Ref.~\cite{A_Terasawa_2019}.

To examine overlap effects, different 
choices of basis sets were 
examined in this paper.
The minimal basis sets 
were
constructed from $3s$, $3p$, $3d$, and $4s$ orbitals for Fe, Co, and Ni atoms, and from $5s$, $5p$, $5d$, and $6s$ orbitals for Nd and Sm atoms.
The $4f$ states of Nd and Sm 
were
treated as spin polarized core states.
When extensive basis sets were adopted, we 
included
orbitals having more spreading basis distributions, which 
resulted 
in larger overlaps between different sites.
To describe the effect of
the
core states, we adopted the fully relativistic pseudopotentials generated by the Morrison-Bylander-Kleinman scheme \cite{MBK}.

\section{Results}\label{sec:results}

\subsection{Detailed Analysis of $J_{ij}$ for bcc Fe}

\begin{figure}[tp]
\begin{center}
\includegraphics[width=72.72mm]{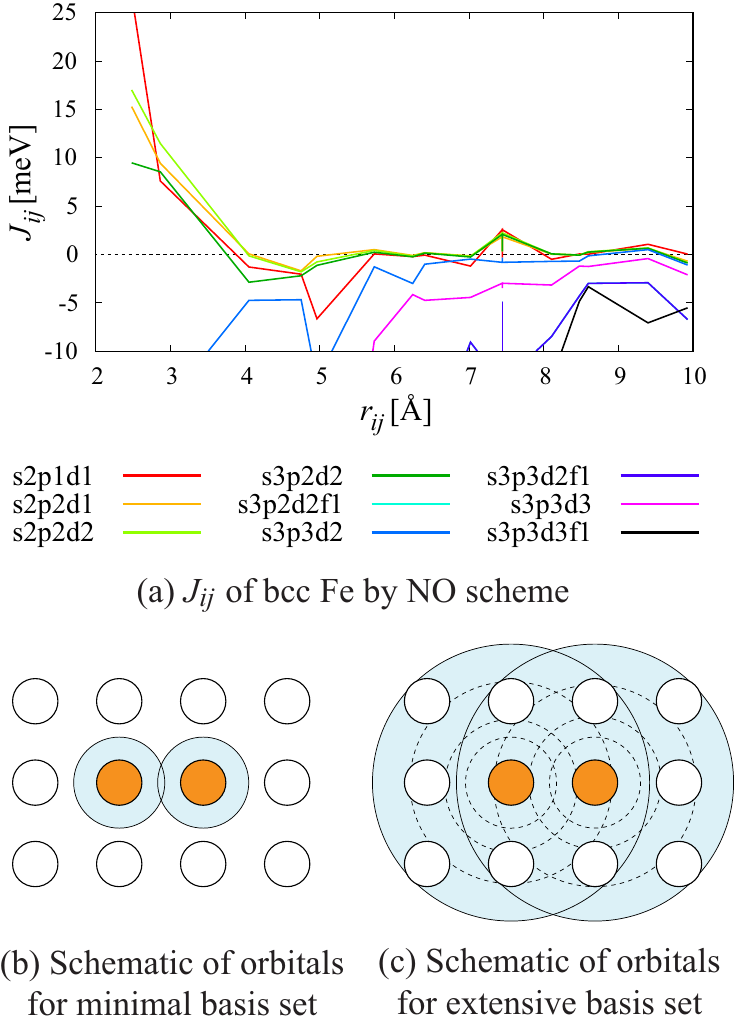}\\
\end{center}
\caption{ 
(a) Exchange coupling constants $J_{ij}$ as functions of atomic 
distance 
$r_{ij}$ for the NO scheme and different choices of basis sets. (b) and (c) Schematics of orbitals for different choices of basis sets. (Color online)
\label{fig:bccFe_1}
}
\end{figure}

We first present the 
exchange coupling constants $J_{ij}$ of bcc Fe in Fig.~\ref{fig:bccFe_1}(a) as functions of atomic distances $r_{ij}$ for the NO scheme. In Fig~\ref{fig:bccFe_1}(a), the lines of different colors correspond to different choices of basis 
sets 
from s2p1d1 
to 
s3p3d3f1.
Here, the notation s$x$p$y$d$z$f$w$ means that the basis set is constructed from $x$ types of $s$ orbitals, $y$ types of $p$ orbitals, $z$ types of $d$ orbitals, and $w$ types of $f$ orbitals. 
The 
total number of basis functions $N_{\mathrm{b}}$ per atomic site 
is thus 
$(x+3y+5z+7w)$ for 
the
notation s$x$p$y$d$z$f$w$.
A remarkable feature can be seen 
in Fig.~\ref{fig:bccFe_1}(a) 
where 
the $J_{ij}$ profiles are similar for s2p2d1, s2p2d2, and s3p2d2 with fluctuations of about a few meV, whereas the $J_{ij}$ profiles deviate strongly to negative values for the larger basis sets.

\begin{figure}[tp]
\begin{center}
\includegraphics[width=70mm]{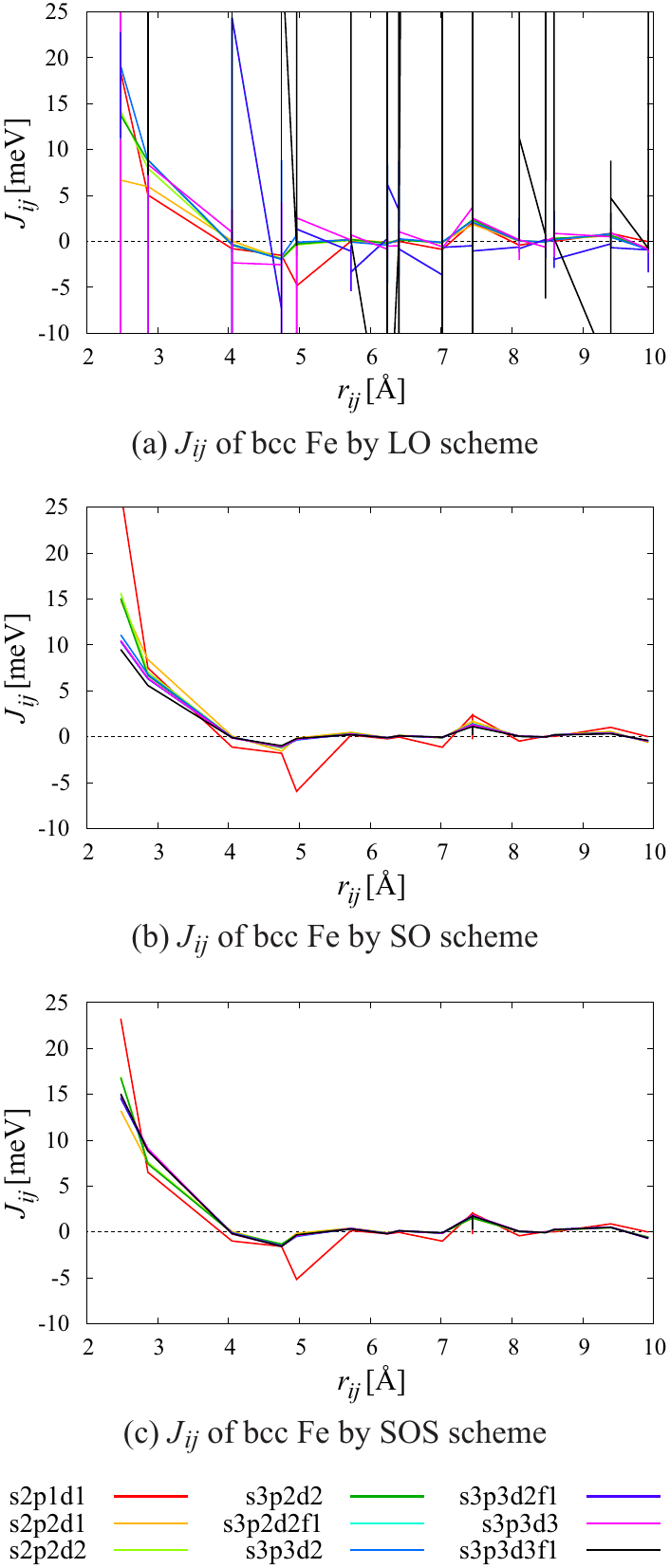}\\
\end{center}
\caption{
Exchange coupling constants $J_{ij}$ of bcc Fe as functions of atomic distances $r_{ij}$ for different orthogonalization schemes and choices of basis sets.
(Color online)
\label{fig:bccFe_2}
}
\end{figure}

This feature seems strange at first sight because the calculations of the electronic states 
themselves 
become more accurate as we increase the total number of basis functions.
To understand the feature, it is necessary to consider the relationship between 
the 
single-site potentials and the LCAO 
Hamiltonian matrix elements.
Figures \ref{fig:bccFe_1}(b) and (c) show the relationships between 
the 
orbitals and atomic sites 
schematically.
When we choose the minimal basis set, we only take into account combinations of orbitals which have small ranges as shown in Fig.~\ref{fig:bccFe_1}(b).
In this case, 
the 
electronic 
potentials 
in the considered ranges can be regarded as effective single-site 
potentials,  
and thus the Liechtenstein formula works well.
When we choose an extensive basis set, in contrast, more spreading basis functions for each site are taken into account in the calculation as shown in Fig.~\ref{fig:bccFe_1}(c).
In this case, the range of a single orbital 
spans 
multiple atoms, and 
the
matrix elements for wide orbitals are no longer 
effective
single-site potentials.
This results in the breakdown of 
the
Liechtenstein formula when 
adopting 
non-orthogonal Hamiltonian for effective potential terms.
While the instability at large $N_{\mathrm{b}}$ 
becomes 
natural with this reasoning, 
the instability results in 
a cumbersome problem: we cannot determine accurate $J_{ij}$ values 
simply 
by adopting NO Hamiltonians as effective single-site potentials 
from the 
LCAO calculation.

In order to stabilize the calculation results for different choices of basis sets, we examined 
the 
two types of orthogonalization schemes which 
have been 
explained in the previous sections, 
namely
the LO and SO schemes, as shown in Figs.~\ref{fig:bccFe_2}(a) and (b).
It is apparent that the LO scheme does not solve the problem 
illustrated
in Fig.~\ref{fig:bccFe_2}(a).
The calculated $J_{ij}$ values do not converge either, and 
fluctuate strongly 
for large $N_{\mathrm{b}}$.
This indicates that the LO scheme fails to represent effective single-site potentials in the presence of overlapped atomic orbitals.
That is, an electron described by an LO function at a specific atomic site feels the potential coming from other atoms, and thus the diagonal elements of 
the 
Hamiltonian in
the 
LO representation include the 
contributions 
of multiple atoms.

In contrast, 
the 
$J_{ij}$ results calculated with the SO scheme show a relatively stable behavior but 
decrease
slightly 
with
increasing $N_{\mathrm{b}}$.
The stability of 
the 
$J_{ij}$ curves in the SO scheme indicates that the 
overlap cancellation
by damped oscillations also works well for the matrix representation of 
the 
Hamiltonian, and 
that
the single-site Hamiltonian is 
well described 
by the SO scheme.

The gradual decrease of $J_{ij}$ in the SO scheme is however non-negligible.
This may 
have resulted 
from the deviation of 
the
spin population depending on the choice of basis set, as pointed out in Ref.~\cite{Y_O_Kvashnin_2015}.
In particular, the SO scheme tends to underestimate the spin population, because it subtracts the components of other atoms from the basis of the focused site.
To eliminate the underestimation, we introduce spin population scaling as follows:
\begin{equation}
J_{ij}^{\mathrm{(SOS)}}
\equiv
\frac{
\Delta n_{i}
}{
\Delta n_{j}^{\mathrm{(SO)}}
}
\frac{
\Delta n_{j}
}{
\Delta n_{j}^{\mathrm{(SO)}}
}
J_{ij}^{\mathrm{(SO)}},
\end{equation}
where $\Delta n_{i}^{\mathrm{(SO)}}$ and $\Delta n_{i}^{\mathrm{(SO)}}$ 
are
the spin population at site $i$ calculated 
using the
NO basis and SO basis, respectively, and $J_{ij}^{\mathrm{(SO)}}$ 
is
the $J_{ij}$ value calculated with the SO scheme.
This spin population scaling is based on the assumption that 
the 
$J_{ij}$ values are proportional to the spin populations at sites $i$ and $j$ for small deviations.
We call this scheme as single-site orthogonalization with spin population scaling (SOS).
The calculated $J_{ij}$ as functions of $r_{ij}$ based on the SOS scheme 
are
shown in Fig.~\ref{fig:bccFe_2}(c).
We can see that the $J_{ij}$ profiles 
converge 
well and 
are 
almost independent of the choice of basis sets for large number of orbitals.

\begin{figure}[tp]
\begin{center}
\includegraphics[width=69.59mm]{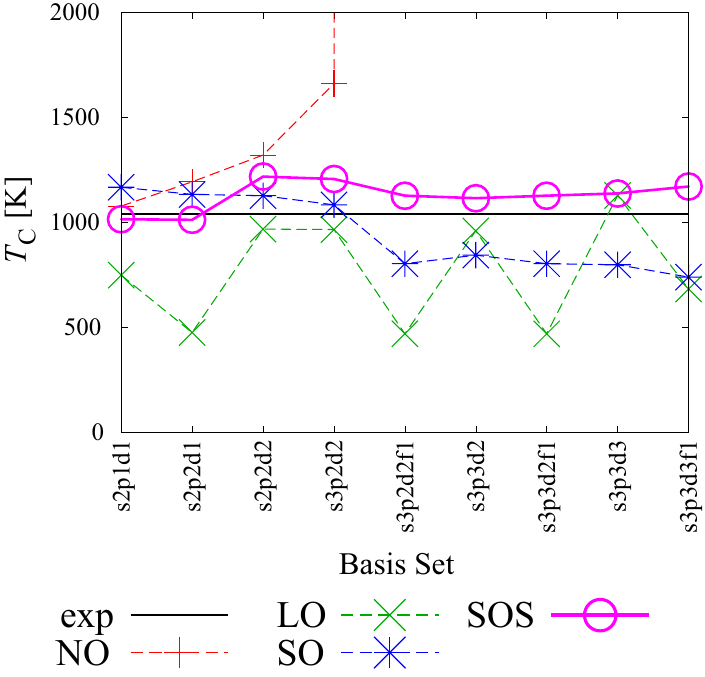}\\
\end{center}
\caption{Calculated Curie temperatures of bcc Fe within mean field approximation for different orthogonalization scheme and different choices of basis sets, together with the experimental values.
(Color online)
\label{fig:fig_bccFe_3}
}
\end{figure}

We also calculated the Curie temperature $T_{\mathrm{C}}$ from 
the 
$J_{ij}$ values.
Within the mean field approximation, the Curie 
temperature 
$T_{\mathrm{C}}$ can be 
obtained 
as the maximum eigenvalue of the matrix $\bm{\Theta}$ 
with 
the matrix elements
\begin{eqnarray}
\big[\bm{\Theta}\big]_{ij}
&=
 &\frac{2}{3 k_{\mathrm{B}}}\times\left\{\begin{array}{ll}
 \sum_{\mathbf{R}\ne\mathbf{0}}J_{\mathbf{0}i,\mathbf{R}i},&i=j\\
 \sum_{\mathbf{R}}J_{\mathbf{0}i,\mathbf{R}j},&i\ne j\\ 
 \end{array}\right.
\label{eq:Theta}
\end{eqnarray}
where $J_{\mathbf{0}i,\mathbf{R}j}$ is the exchange coupling constant between site $i$ at cell $\mathbf{0}$ and site $j$ at cell $\mathbf{R}$.
The summation in Eq.~(\ref{eq:Theta}) was derived by subtracting the self-interaction term from the periodic sum as described in Ref.~\cite{A_Terasawa_2019}.
The calculated $T_{\mathrm{C}}$ for different basis sets are shown in Fig.~\ref{fig:fig_bccFe_3}.
As expected from the $J_{ij}$ profiles, the $T_{\mathrm{C}}$ calculated with the SOS scheme shows convergent behavior as $N_{\mathrm{b}}$ increases and converges 
at a value about a few tens of percent higher than the experimental value, while that derived from the SO scheme gradually decreases with increasing $N_{\mathrm{b}}$.
In contrast, the $T_{\mathrm{C}}$s calculated from the NO and LO schemes show large deviations at large $N_{\mathrm{b}}$.
These results show that the NO and LO schemes are unsuitable for the Liechtenstein calculation within the LCAO approximation, and that the SOS scheme is the most stable scheme for calculating $T_{\mathrm{C}}$ among the schemes examined.

\subsection{Curie temperatures of hcp Co and fcc Ni}

\begin{figure}[tp]
\begin{center}
\includegraphics[width=69.59mm]{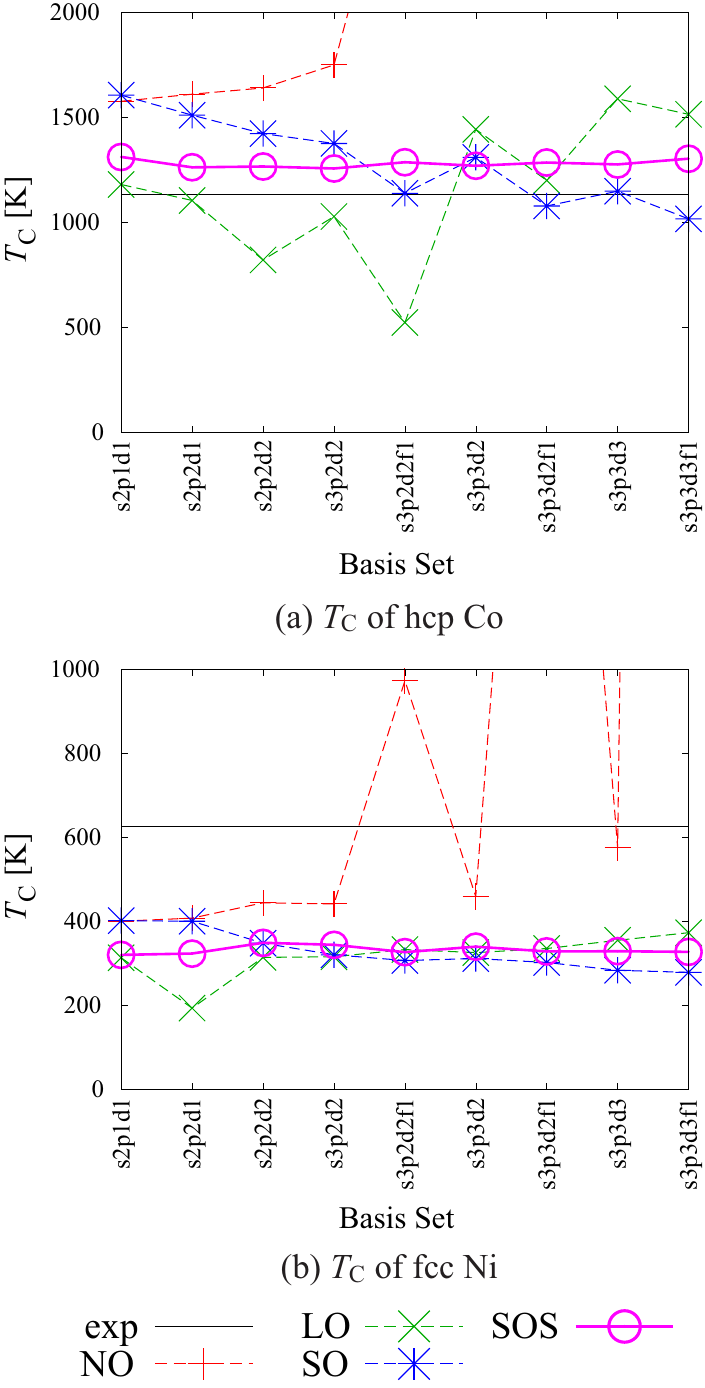}\\
\end{center}
\caption{Calculated Curie temperatures of transition metals within mean field approximation for different orthogonalization 
schemes
and different choices of basis sets, together with the experimental values.
(Color online)
\label{fig:Co_Ni}
}
\end{figure}

In addition to bcc Fe, we calculated $J_{ij}$ and $T_{\mathrm{C}}$ of hcp Co and fcc Ni for different orthogonalization schemes and different choices of basis sets.
Figure \ref{fig:Co_Ni} shows the calculated Curie temperatures of hcp Co and fcc Ni.
It can be seen in Fig.~\ref{fig:Co_Ni} that the Curie temperatures derived from the SOS scheme show 
convergent 
behavior while those derived from the SO scheme gradually 
decrease.
The results for the NO and LO schemes show unstable behavior except for the LO results for fcc Ni.
These results are another evidence 
for 
the SOS scheme 
being 
the most stable scheme for the Liechtenstein calculation.

\begin{figure*}[tp]
\begin{center}
\includegraphics[width=143.71mm]{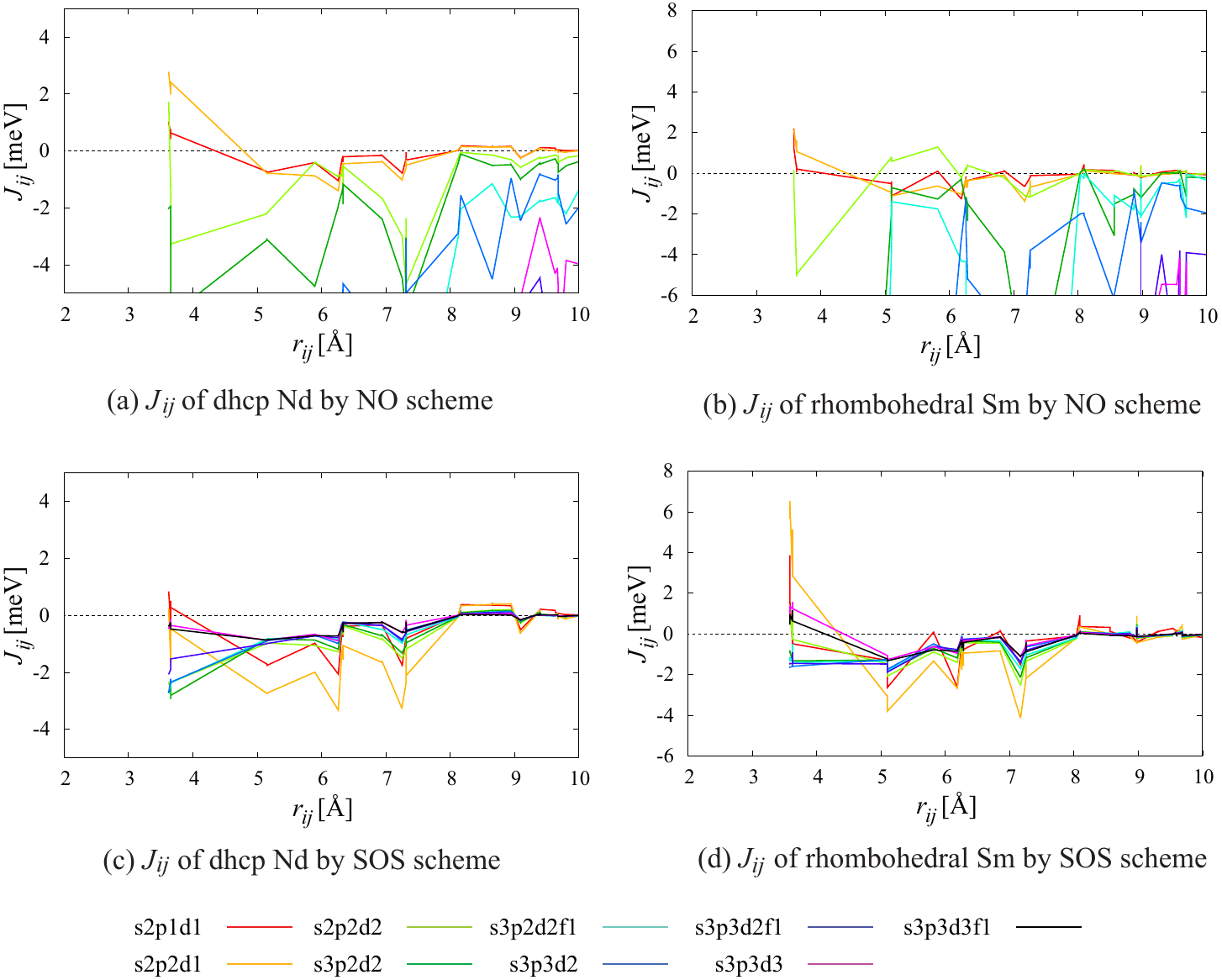}\\
\end{center}
\caption{
Exchange coupling constants $J_{ij}$ of (a) and (b) dhcp Nd and (c) and (d) rhombohedral Sm as functions of atomic distances $r_{ij}$ 
calculated
by the SOS scheme and different choices of basis sets.
(Color online)
\label{fig:Nd_Sm}
}
\end{figure*}

\subsection{$J_{ij}$ profiles for dhcp Nd and rhombohedral Sm}

Next, we calculated $J_{ij}$ for dhcp Nd and rhombohedral Sm 
with 
different orthogonalization schemes and choices of basis sets.
Figure \ref{fig:Nd_Sm}(a)-(b) shows the calculated $J_{ij}$ as functions of atomic distance $r_{ij}$ for the NO scheme, and Figure \ref{fig:Nd_Sm}(c)-(d) for the SOS scheme.
It 
can be seen 
in Figure \ref{fig:Nd_Sm}(a)
and 
(b) that the calculated $J_{ij}$ values by the NO scheme 
vary 
more drastically 
with the 
basis sets than those of transition metals in the previous section, and 
that 
it is almost impossible to determine the correct results 
from
the NO calculations.
In contrast, the $J_{ij}$ values 
from
the SOS scheme 
converge 
slowly 
with increasing 
basis set 
size,
as 
can be
seen in Fig.~\ref{fig:Nd_Sm}(c) and (d).
Moreover, 
small 
negative $J_{ij}$ values for long 
ranges 
are obtained for both cases in the SOS calculations.
These results 
agree 
with experimental reports 
on 
Nd and Sm 
exhibiting 
spiral and complicated magnetic orders \cite{Coeybook}, which are attributed to antiferromagnetic exchange interactions.

\section{Summary}
We introduced a new orthogonalization scheme called single-site orthogonalization (SO)
for calculating exchange coupling constants $J_{ij}$ within the LCAO approximation, and 
compared the calculated $J_{ij}$ for bcc Fe, hcp Co, fcc Ni dhcp Nd and rhombohedral Sm by the SO scheme with those calculated by the non-orthogonal (NO) and L\"owdin orthogonalization (LO) schemes.
We found
that the SO scheme underestimates $J_{ij}$ slightly as the number of basis functions 
$N_{\mathrm{b}}$ 
increases,
whereas the NO and LO schemes give strongly 
fluctuating 
results for large $N_{\mathrm{b}}$.
The underestimation by the SO scheme can be 
well corrected 
by introducing spin-population scaling, which we call single-site orthogonalization with spin population scaling (SOS).
Using the SOS scheme, we 
successfully obtained 
converged Curie temperatures for transition metals and small negative $J_{ij}$ 
values
for rare earth metals.
We believe that the formalism introduced in this study opens up new 
prospects 
for understanding the coexistence of localized and itinerant electrons.

\begin{acknowledgments}
The authors 
thank
Hisazumi Akai and Munehisa Matsumoto for fruitful discussions and valuable comments.

This work was supported in part by MEXT, Japan as Program for Promoting Researches on the Supercomputer Fugaku, DPMSD, the Elements Strategy Initiative Project (ESICMM, Grant No.~JPMXP0112101004) under the auspices of MEXT, as well as KAKENHI Grant No.~17K04978. 
Some of the calculations were performed using the supercomputers at ISSP, The University of Tokyo, and TSUBAME, Tokyo Institute of Technology, as well as the K computer, RIKEN Project Nos.~hp180206, and hp190169).
\end{acknowledgments}

\end{document}